\documentstyle[11pt,epsf,float]{article}
\newcounter{subequation}[equation]

\makeatletter

\def\thesubequation{\theequation\@alph\c@subequation}
\def\@subeqnnum{{\rm (\thesubequation)}}
\def\slabel#1{\@bsphack\if@filesw {\let\thepage\relax
   \xdef\@gtempa{\write\@auxout{\string
      \newlabel{#1}{{\thesubequation}{\thepage}}}}}\@gtempa
   \if@nobreak \ifvmode\nobreak\fi\fi\fi\@esphack}
\def\subeqnarray{\stepcounter{equation}
\let\@currentlabel=\theequation\global\c@subequation\@ne
\global\@eqnswtrue
\global\@eqcnt\z@\tabskip\@centering\let\\=\@subeqncr
$$\halign to \displaywidth\bgroup\@eqnsel\hskip\@centering
  $\displaystyle\tabskip\z@{##}$&\global\@eqcnt\@ne
  \hskip 2\arraycolsep \hfil${##}$\hfil
  &\global\@eqcnt\tw@ \hskip 2\arraycolsep
  $\displaystyle\tabskip\z@{##}$\hfil
   \tabskip\@centering&\llap{##}\tabskip\z@\cr}
\def\endsubeqnarray{\@@subeqncr\egroup
                     $$\global\@ignoretrue}
\def\@subeqncr{{\ifnum0=`}\fi\@ifstar{\global\@eqpen\@M
    \@ysubeqncr}{\global\@eqpen\interdisplaylinepenalty \@ysubeqncr}}
\def\@ysubeqncr{\@ifnextchar [{\@xsubeqncr}{\@xsubeqncr[\z@]}}
\def\@xsubeqncr[#1]{\ifnum0=`{\fi}\@@subeqncr
   \noalign{\penalty\@eqpen\vskip\jot\vskip #1\relax}}
\def\@@subeqncr{\let\@tempa\relax
    \ifcase\@eqcnt \def\@tempa{& & &}\or \def\@tempa{& &}
      \else \def\@tempa{&}\fi
     \@tempa \if@eqnsw\@subeqnnum\refstepcounter{subequation}\fi
     \global\@eqnswtrue\global\@eqcnt\z@\cr}
\let\@ssubeqncr=\@subeqncr
\@namedef{subeqnarray*}{\def\@subeqncr{\nonumber\@ssubeqncr}\subeqnarray}
\@namedef{endsubeqnarray*}{\global\advance\c@equation\m@ne%
                           \nonumber\endsubeqnarray}

\makeatletter
\@addtoreset{equation}{section}
\makeatother
\renewcommand{\theequation}{\thesection.\arabic{equation}}

\def\dalemb#1#2{{\vbox{\hrule height .#2pt
        \hbox{\vrule width.#2pt height#1pt \kern#1pt
                \vrule width.#2pt}
        \hrule height.#2pt}}}

\let\a=\alpha

 \let\t=\tau   \let\f=\phi
  
\let\w=\omega  \let\D=\Delta

\let\la=\label  
\let\se=\section  
\def\nn{\nonumber} \def\bd{\begin{document}} \def\ed{\end{document}}
\def\ds{\documentstyle} \let\fr=\frac \let\bl=\bigl \let\br=\bigr
\let\Br=\Bigr \let\Bl=\Bigl
\let\bm=\bibitem
\let\na=\nabla
\let\pa=\partial \let\ov=\overline
\def\ie{{\it i.e.\ }}
\def\wlg{{\it w.l.o.g.\ }}
\newcommand{\be}{\begin{equation}}
\newcommand{\ee}{\end{equation}}
\def\ba{\begin{array}}
\def\ea{\end{array}}
\def\ft#1#2{{\textstyle{{\scriptstyle #1}\over {\scriptstyle #2}}}}
\def\fft#1#2{{#1 \over #2}}
\def\del{\partial}
\def\sst#1{{\scriptscriptstyle #1}}
\def\oneone{\rlap 1\mkern4mu{\rm l}}
\def\e7{E_{7(+7)}}
\def\td{\tilde}
\def\wtd{\widetilde}
\def\im{{\rm i}}
\def\bog{Bogomol'nyi\ }
\def\tq{{\tilde q}}
\def\hast{{\hat\ast}}
\def\0{{\sst{(0)}}}
\def\1{{\sst{(1)}}}
\def\2{{\sst{(2)}}}
\def\3{{\sst{(3)}}}
\def\4{{\sst{(4)}}}
\def\5{{\sst{(5)}}}
\def\6{{\sst{(6)}}}
\def\7{{\sst{(7)}}}
\def\8{{\sst{(8)}}}
\def\ne{{\sst{(n)}}}
\def\oo{{\"o}}
\def\hA{\hat{\cal A}}
\def\ns{{\sst {\rm NS}}}
\def\rr{{\sst {\rm RR}}}
\def\tH{{H_A}}
\def\htH{{\hat{H}_A}}
\def\ctK{{\widetilde {\cal K}}}
\def\tB{{\widetilde B}}
\def\tw{{\w_A}}
\def\cF{{\cal F}}
\def\tF{{\wtd F}}
\def\v{{{\cal V}}}
\def\Z{\rlap{\sf Z}\mkern3mu{\sf Z}}
\def\ep{{\varepsilon}}
\def\IIA{{\rm IIA}}
\def\IIB{{\rm IIB}}
\def\ads{{\rm AdS}}
\def\R{\rlap{\rm I}\mkern3mu{\rm R}}
\def\vp{{\varphi}}
\def\ns{{\sst{\rm NS}}}
\def\rr{{\sst{\rm RR}}}
\def\cF{{\cal F}}
\def\cA{{\cal A}}
\def\cB{{\cal B}}
\def\cP{{\cal P}}
\def\cQ{{\cal Q}}
\def\cK{{\cal K}}
\def\cT{{\cal T}}
\def\cM{{\cal M}}
\def\cJ{{\cal J}}
\def\cG{{\cal G}}
\def\hA{{\hat{\cal A}}}
\def\td{\tilde} \def\wtd{\widetilde}
\def\Z{\rlap{\sf Z}\mkern3mu{\sf Z}}
\def\hhs{{\qquad}}
\def\hs{\,\,}
\def\wdg{\wedge}
\def\all{\forall}
\def\hT{\hat{T}}
\def\hV{\hat{V}}
\def\hL{\hat{L}}
\def\hX{\hat{X}}
\def\hY{\hat{Y}}
\def\hH{\hat{H}}

\def\t1{e_{_T}}
\def\v1{e_{_V}}
\def\ct{e_{_{TTV}}}
\def\cv{e_{_{VTV}}}
\def\tt{e_{_{TTTTV}}}
\def\tv{e_{_{VTTTV}}}
\def\vt{e_{_{TTVTV}}}
\def\vv{e_{_{VTVTV}}}

\newcommand{\ho}[1]{$\, ^{#1}$}
\newcommand{\hoch}[1]{$\, ^{#1}$}
\newcommand{\bea}{\begin{eqnarray}}
\newcommand{\eea}{\end{eqnarray}}
\newcommand{\ra}{\rightarrow}
\newcommand{\Ra}{\Rightarrow}
\newcommand{\Lra}{\Longrightarrow}
\newcommand{\lra}{\longrightarrow}
\newcommand{\LRa}{\Leftrightarrow}
\newcommand{\ap}{\alpha^\prime}
\newcommand{\bp}{\tilde \beta^\prime}
\newcommand{\tr}{{\rm tr} }
\newcommand{\Tr}{{\rm Tr} }
\newcommand{\NP}{Nucl. Phys. }
\newcommand{\miphys}{\it George P. and Cynthia W. Mitchell Institute 
for Fundamental Physics,\\
Texas A\&M University, College Station, TX 77843-4242}
\newcommand{\tamphys}{\it Center for Theoretical Physics,
Texas A\&M University, College Station, TX 77843-4242}
\newcommand{\upenn}{\it Dept. of Physics and Astronomy,
University of Pennsylvania,
Philadelphia, PA 19104}
\newcommand{\auth}{Sante R. Scuro and Siu A. Chin }

\begin{document}

\begin{flushright}
\hfill{MIFP-04-23}\\
\hfill{\bf math-ph/0411086}\\
\hfill{November 2004}\\
\end{flushright}


\begin{center}
{\large {\bf Forward Symplectic Integrators and \\
the Long Time Phase Error in Periodic Motions}}

\vspace{20pt}

\auth

\vspace{10pt}
{\hoch{}\miphys}

\vspace{30pt}
\underline{ABSTRACT}
\end{center}

   We show that when time-reversible symplectic algorithms are used to
solve periodic motions, the energy error after one period is generally
two orders higher than that of the algorithm.
By use of correctable algorithms, we show that the
phase error can also be eliminated two orders higher
than that of the integrator. The use of fourth order forward
time step integrators can result in sixth order accuracy for the phase
error and eighth accuracy in the periodic energy. We study the
1-D harmonic oscillator and the 2-D Kepler problem in great details,
and compare the effectiveness of some recent fourth order algorithms.

{\vfill\leftline{}\vfill
\vskip 10pt \footnoterule
{\footnotesize\hoch{} Research supported in part by NSF grant
DMS-0310580\vskip -12pt}  \vskip  14pt}

\pagebreak
\setcounter{page}{1}

\tableofcontents
\addtocontents{toc}{\protect\setcounter{tocdepth}{2}}
\newpage

\section{Introduction}

    Symplectic integrators\cite{wisdom,yoshi,mcl95,cha96,mcl02} preserve
Poincar\'e invariants when integrating classical trajectories.
For periodic motion, their energy errors are bounded
and periodic, in contrast to non-symplectic Runge-Kutta type
algorithms\cite{bat99} whose energy error grows linearly
with the number of periods\cite{shita,gladman,cano}. Energy conservation
alone suggests that symplectic algorithms are better long time integrator
of classical motions. However, for periodic motion, even symplectic
algorithms are not immune from the linear growth of the phase error
\cite{shita,gladman,cano}. Whereas the energy error is the error
of the {\it action} variable, the phase error is the error of the
{\it angle} variable. Of the two, the phase error is even more important in
determining the long term accuracy of trajectories. For example, when
symplectic algorithms are used to compute the Keplerian orbit,
the elliptical orbit is easily seen to precess. The precession
is of nearly constant radius. Since the semi-major axis of the ellipse is
fixed by the initial energy, the constancy of the precession radius implies
excellent energy conservation. Yet in spite of that, the precession itself
implies that the trajectory is highly inaccurate. This orbital precession is a
direct manifestation of phase error. Thus to preserve the long term
accuracy of periodic trajectories, despite the primacy of energy
conservation\cite{ge}, one must seek to reduce the phase error directly.

    For periodic motion, the only error that matters is error that persists 
after one period\cite{cano}. A fundamental finding of this work is that,
for periodic motion after one period, the energy error is at least $(\Delta t)^2$
times that of the phase error, where $\Delta t$ is the time step size
used. Thus at small $\Delta t$ the phase error is the dominant
error governing the long term accuracy of periodic motion. Moreover, we
show that the phase error of the symplectic 
corrector\cite{wis96,mcl962,mar96,mar97,blan99} kernel algorithm
is $(\Delta t)^2$ times the phase error of other algorithms nominally
of the same order. Recently, one of
us\cite{symcorr} has made explicit the ``correctability" requirement in
deriving a correctable kernel algorithm. This criterion determines
the optimal symplectic algorithms for solving periodic
motion. The corrector algorithm has its origin in
canonical perturbation theory\cite{suss}. It has been
studied extensively\cite{wis96,mcl962,mar96,mar97,blan99} for its
labor saving feature of only having to iterate the kernel algorithm.
Here we draw the connection between symplectic corrector
algorithms and the phase error in periodic motion. Much of our analysis
is analytical rather than numerical, so that one can understand the result
in a transparent way. We also found that {\it forward} time step symplectic
algorithms\cite{suzfour,chin,hiord,chinchen02,chinchen03} generally
have much smaller phase errors than traditional algorithms
with backward intermediate time steps\cite{mcl95,mcl02,forest,ome02,ome03}.

     In this work, we will analyze in detail the two fundamental prototypes
of periodic motion: the 1-D harmonic oscillator and the 2-D Kepler orbit.
We are not interested in solving the harmonic oscillator per se, 
but only in using it as a vehicle for understanding the phase error 
and the working of our algorithms. It is only with such a simple model 
that we can show analytically how the phase error can be reduced by fine tuning 
the algorithm. To the extent that harmonic motion is the simplest periodic
motion, this is clearly a necessary first step for proposing any scheme
of phase error reduction. In the 2-D Kepler case,
we demonstrate the usefulness of forward symplectic algorithms as compared to
existing negative time step algorithms. For completeness, we begin with a brief
review of the operator construction of symplectic algorithms, followed by
a synopsis of symplectic corrector algorithms. In section \ref{ho2nd}, we
illustrate the basic idea of our analysis by showing how a second order
algorithm can achieve fourth order accuracy in the phase error when solving the
1-D harmonic oscillator. In section \ref{ho4th}, we repeat the same analysis
for a class of fourth order forward algorithms.
Error terms up to eighth order are computed by use of the Lie series\cite{dragt}
expansion. Beyond eighth order, the error terms can be determined by 
exactly solving the matrix model. All these are done analytically.
We repeat the analysis for the Kepler problem in section \ref{kepl}.
Here, we compare the phase error numerically for a number of
recent fourth order symplectic algorithms. We summarize our conclusions
in section \ref{conclusion}. For the reader's convenience, some lengthy
formulae and explicit calculations are given in the Appendix.

\section{Operator Factorization}\la{opfa}

    Symplectic algorithms can be derived most simply on the basis of operator
factorization. (See the excellent review by
Yoshida\cite{yoshi} and earlier references therein.)
For any dynamical variable $W(q_i,p_i)$,
its time evolution is given by the Poisson bracket, and therefore by the
corresponding Lie operator $\hat H$ associated with the Hamiltonian
function $H(q_i,p_i)$, \ie
\bea
\fft{dW}{dt} =\{W,H\}
        &\equiv&\fft{\pa W}{\pa q_i}\, \fft{\pa H}{\pa p_i}
         -\fft{\pa W}{\pa p_i}\, \fft{\pa H}{\pa q_i}\, ,\la{posbk} \\
             &=& \Bigl(\fft{\pa H}{\pa p_i}\,\fft{\pa}{\pa q_i}
         -\fft{\pa H}{\pa q_i}\,\fft{\pa}{\pa p_i}\Bigr)W
             = \hH\, W\, . \la{liou}
\eea
(Repeated indices imply summation). More generally, for any dynamical
variable $Q$, we can define its associated Lie operator
$\hat Q$ via the Poisson bracket
\be
\hat{Q}\, W=\{W,Q\}\, .
\la{liop}
\ee
As we will see, this fundamental operator mapping
underpins the entire development of symplectic integrators.

The operator equation (\ref{liou}) can be formally solved via
\be
W(t)={\rm e}^{t\, \hH}W(0)\, .
\la{formsol}
\ee
Symplectic algorithms are derived by approximating the evolution
operator ${\rm e}^{t\, \hH}$ for a short time in a product form.
For Hamiltonian function of the standard separable form,
\be
H({\bf q},{\bf p}) = T({\bf p})+V({\bf q}),\qquad {\rm with}\qquad
T({\bf p}) = \fft{1}{2}p_i p_i\, ,\la{ham}
\ee
the Hamiltonian operator (\ref{liou}) is also separable,
\be
\hH=\hT+\hV\, , \la{htv}
\ee
with first order differential operators $\hat{T}$ and $\hat{V}$
given by
\be
\hT \equiv\,\,  \fft{\pa T}{\pa p_i}\fft{\pa}{\pa q_i}
=p_i \fft{\pa}{\pa q_i}\, ,
 \la{htop}
 \ee
 \be
\hV \equiv -\fft{\pa V}{\pa q_i} \fft{\pa}{\pa p_i}
=F_i({\bf q}) \fft{\pa}{\pa p_i}\, .
\la{hvop}
\ee
Note that $\hH$, $\hT$ and $\hV$ individually satisfy the defining
equality (\ref{liop}).

The corresponding Lie transforms\cite{dragt}
${\rm e}^{\,\ep\, \hT}$ and ${\rm e}^{\,\ep\, \hV}$,
are then displacement operators which shift $q_i$ and $p_i$ forward in
time via
\be
{\bf q}\ra {\bf q}+\ep\, {\bf p}
\qquad{\rm and}\qquad {\bf p}\ra {\bf p}+\ep\, {\bf F}\, .\la{pqsh}
\ee
Thus, if ${\rm e}^{\ep\hH}$ can be factorized into products of
Lie transforms ${\rm e}^{\ep\hT}$ and
${\rm e}^{\ep\hV}$, then each factorization gives rise to an integrator
for evolving the system forward in time. Most of the existing literature
on symplectic algorithms is concerned with decomposing ${\rm e}^{\ep\hH}$
to arbitrarily higher order in the product form of
\be
{\rm e}^{\ep(\hT+\hV)}\approx \prod_{i=1}^N
{\rm e}^{t_i\ep\hT}{\rm e}^{v_i\ep\hV}, \la{prod}
\ee
with a well chosen set of factorization coefficients
$\{t_i,v_i\}$. In most cases, we will consider only the
left-right symmetric factorization schemes
such that  either $t_1=0$ and
$v_i=v_{N-i+1}$, $t_{i+1}=t_{N-i+1}$, or $v_N=0$ and
$v_i=v_{N-i}$, $t_{i}=t_{N-i+1}$. In either cases, the algorithm
is exactly time-reversible, and the energy error terms
can only be an even function of $\ep$.
Such a symmetric factorizations is then at least second
order. As first proved by Sheng\cite{sheng}, and
Suzuki\cite{suzukinogo}, beyond second order, decompositions of
the form (\ref{prod}) must contain some negative coefficients
$t_i$ and $v_i$. Goldman and Kaper\cite{goldman} further proved
that beyond second order, there must be at least be one pair of
negative coefficients $(t_i,v_i)$. To circumvent this backward
time step restriction\cite{suzfour,chin}, one must factorize the
evolution operator in terms of operators $\hat T$, $\hat V$ {\it
and} the commutator $[\hat{V},[\hat{T},\hat{V}]]$. In this work,
we will further demonstrate that these forward symplectic algorithms
are also effective in reducing the phase error.

\se{Symplectic Corrector Algorithms}\la{symcor}
To see the relevance of symplectic corrector algorithms to periodic motion,
we recapitulate some recent results\cite{symcorr}.
Let $\cT_A$ be a symmetric, approximate factorization of the short time
evolution operator ${\rm e}^{\,\ep(\hT+\hV)}$,
\be
\cT_A=\prod_{i=1}^N
{\rm e}^{t_i\ep \hT}{\rm e}^{v_i\ep \hV}={\rm e}^{\,\ep\hat H_A}\, ,
\la{arho}
\ee
then the approximate Hamiltonian operator $\hat H_A$ must be even in
$\ep$, \ie
\be
\hat H_A=\hT+\hV+\ep^2(\, e_{TTV}[\hT,[\hT,\hV]]+e_{VTV}[\hV,[\hT,\hV]]\,)
+O(\ep^4)\, ,\la{ha}
\ee
with error coefficients $e_{TTV}$, $e_{VTV}$ determined by
factorization coefficients $\{t_i,v_i\}$. Consider the similarity
transformed propagator,
\be
\cT_A^\prime=S\cT_A S^{-1}=S{\rm e}^{\,\ep \hat H_A}S^{-1}
={\rm e}^{\,\ep( S \hat H_A S^{-1})}={\rm e}^{\,\ep\hat H_A^\prime}\, ,
\label{tha}
\ee
where the last equality defines the transformed
Hamiltonian $\hat H_A^\prime$.
If now we take
\be
S=\exp[\ep \hat C]\la{scor}\, ,
\ee
where $\hat C$ is the corrector, then the following fundamental result
\be
\hat H_A^\prime=e^{\ep \hat C}\hat H_A{\rm e}^{-\ep \hat C}
=\hat H_A+\ep[\hat C,\hat H_A]
+{1\over 2!}\ep^2[\hat C,[\hat C,\hat H_A]]+\cdots\, ,
\la{fcor}
\ee
implies that
\be
\hat H_A^\prime=
\hT+\hV+\ep^2(\, e_{TTV}[\hT,[\hT,\hV]]+e_{VTV}[\hV,[\hT,\hV]]\,)
+\ep[\hat C,\hT+\hV]+ \cdots\, .
\la{gcor}
\ee
One immediately sees that the choice
\be
\hat C=\ep\,\, c_{TV}[\hT,\hV]\, ,\la{cchoi}
\ee
would eliminate either second order error term with $c_{TV}=e_{TTV}$ or
$c_{TV}=e_{VTV}$. More importantly, if $\hat H_A$ is constructed such that
\be
e_{TTV}=e_{VTV}\, , \la{eqcof}
\ee
then {\it both} error terms can be eliminated by the corrector.
Thus for such an approximate $\cT_A$, the transformed propagator
$\cT_A^\prime$ will be fourth order. This is the fundamental
``correctability" requirement for correcting a second order $\cT_A$
to fourth order\cite{symcorr}. In general, the corrector can be more
complicated than the kernel algorithm $\cT_A$. However, when one iterates
$\cT_A^\prime$, all intermediate correctors cancel and only the initial
and final corrector remains. For periodic motion, even the initial and
the final corrector would have cancelled after exactly one period.
Hence even if $\cT_A$ is only second order, if it satisfies the
correctability requirement (\ref{eqcof}), then its error
after exactly one period would be fourth order! Thus among all second order
algorithms, those that are ``correctable", \ie satisfy the the
correctability requirement (\ref{eqcof}), would be two orders better.
With a correctable algorithm, we will show later that the phase
error is improved intrinsically even without applying the corrector. 
However, if the step size $\ep$ is not commensurate with the period,
one may step-over the minimum of the error function without knowing
that it is there. In this case, it is essential to apply the corrector 
just prior to computing any observable. The advantage of
a corrector algorithm is that for long-time integration, one usually
only needs to apply the corrector sparingly at a few selected points
in time. 

This correctability requirement can be generalized to higher order.
At higher orders, $\hat H_A$ will have error terms of the form
$[\hat T,\hat Q_i]$ and $[\hat V,\hat Q_i]$ where $\hat Q_i$ are some
higher order commutator generated by $\hat T$ and $\hat V$. If $\hat H_A$
is of order $2n$ in $\ep$, then $H_A^\prime$ can be of order $2n+2$ if
$\hat H_A$'s error coefficients
for $[\hat T,\hat Q_i]$ and $[\hat V,\hat Q_i]$ are {\it equal} for
each $\hat Q_i$.
This is the fundamental corrector insight of \cite{symcorr}.
In the following sections, we will demonstrate how this insight can be
used to reduce the phase error in practical applications.

\section{The Modified Hamiltonian and Error Structure}\la{commcor}
The distinct advantage of symplectic algorithms is not only that they
preserve all Poincar\'e invariants, but that their corresponding
modified Hamiltonians and error structures can be systematically determined.
This is of paramount importance when one seeks to understand the
fundamental cause of an algorithm's error. To illustrate the approach,
we begin by analyzing the simplest, first order factorization,
\be
{\rm e}^{\ep\hT}{\rm e}^{\ep\hV} = {\rm e}^{\ep\htH}
\, ,\la{exprod}
\ee
where $\htH$ is the approximate Hamiltonian operator
\be
\htH= \hH+\fft12\ep[\hT,\hV]+\fft1{12}\ep[\hT,[\hT,\hV]]
   -\fft1{12}\ep[\hV,[\hT,\hV]]+\dots\, . \la{bch}
\ee
of the algorithm. This follows directly from
Baker-Campbell-Hausdorff (BCH) formula. Thus the algorithm evolves the
system according to the modified Hamiltonian $\htH$ rather than the
original Hamiltonian $\hat{H}$.
Nevertheless, the Hamiltonian structure of the system is preserved.
As $\ep\rightarrow 0$, one recovers the original dynamics. Moreover,
knowing $\htH$ allows us to determine the actual Hamiltonian function
$\tH$ which governs the algorithm's evolution. This can be done
systematically by use of the Lie-Poisson bracket correspondence.
To make this part of the discussion self-contained,
we briefly summarize some pertinent results.

    From the fundamental defining equality (\ref{liop}), we can deduce
$\tH$ via
\be
\htH\, W =\{W,\tH\}\, ,
\la{happrox}
\ee
if we know how commutators of $\hT$ and $\hV$ transform back into
functions under the operator mapping (\ref{liop}).
By repeated applications of (\ref{liop}), we have
\bea
[\hT,\hV]\,\,W
&=&\hT\{W,V\}-\hV\{W,T\}\, ,\nn\\
&=&\{\{W,V\},T\}-\{\{W,T\},V\}\, ,\nn\\
&=&\{W,\{V,T\}\}\, ,
\la{comeq}
\eea
where the last equality follows from the
Jacobi identity
$$
\{\{W,V\},T\}+\{\{T,W\},V\}+\{\{V,T\},W\}=0\, .
$$
Equality (\ref{comeq}) implies the following correspondence
between commutators of Lie operators and
Poisson brackets of dynamical variables:
\be
[\hT,\hV] \lra \{V,T\}=-\{T,V\}\, .\la{corr2}
\ee
There is thus a order reversal, or a simple sign change,
in going from Lie commutators to Poisson brackets.
(There is no such order reversal in the usual
correspondence between quantum mechanical commutators and
Poisson brackets.) This order reversal will only
change the sign of odd-order brackets, as illustrated in
the following examples:
\bea
{[\hV,[\hT,\hV]]} &\lra & \{\{V,T\},V\}=\{V,\{T,V\}\}\, ,\la{corr3} \\
{[\hT,[\hV,[\hT,\hV]]]} &\lra&
\{\{\{V,T\}\, ,V\}\, ,T\}=
-\{T,\{V,\{T,V\}\}\}\, .
\nn
\eea
Applying this to (\ref{happrox}) gives, term by term,
\bea
\htH\,W &=& \hH\,W+\fft12\,\ep\,[\hT,\hV]\,W
+\fft1{12}\,\ep^2\,[\hT,[\hT,\hV]]\,W
-\fft1{12}\,\ep^2\,[\hV,[\hT,\hV]]\,W+\dots\, ,\nn \\
\{W,\tH\} &=& \{W,H\}+\{W,\fft12\,\ep\,\{V,T\}\}
+\{W,\fft1{12}\,\ep^2\, \{\{V,T\}\,,T\}\}
-\dots\, ,\la{corrh}
\eea
from which we can identify,
\be
\tH = H-\fft12\,\ep\,\{T,V\}+\fft1{12}\,\ep^2\,\{T,\{T,V\}\}
-\fft1{12}\,\ep^2\,\{V,\{T,V\}\}+\dots\, .
\la{happ}
\ee
This general result merely transcribe expressions of Lie
commutators into Poisson brackets. It is valid regardless
of the form of the Hamiltonian. For the separable
Hamiltonian (\ref{ham}), we have specific results
\be
\{T,V\}=-\fft{\pa T}{\pa p_j}\fft{\pa V}{\pa q_j}\equiv -p_{j}V_{j}\, ,
\label{ptv}
\ee
\be
\{T,\{T,V\}\}=-\fft{\pa T}{\pa p_i}\fft{\pa \{T,V\}}{\pa q_i}
=p_{i}V_{ij}p_{j}\, ,
\la{pttv}
\ee
\be
\{V,\{T,V\}\}=\fft{\pa V}{\pa q_i}\fft{\pa \{T,V\}}{\pa p_i}
=-V_{i}V_{i}\, .
\la{pvtv}
\ee
Since $T=T(\{p_i\})$ and $V=V(\{q_i\})$, there is no ambiguity
about the meaning of subscripts on $T_i$ or $V_j$. Also,
since $T_{ij}=\delta_{ij}$, we therefore have,
\be
\tH = H+\fft12\,\ep\,p_{i}\,V_{i}
+\fft1{12}\,\ep^2\,p_{i}V_{ij}p_{j}
+\fft1{12}\,\ep^2\,V_{i}V_{i}+\dots\, .\la{power}
\ee
In general, the algorithm's approximate Hamiltonian is
non-separable and more complicated than the original Hamiltonian.
Similar expression has been given by Yoshida\cite{yoshi} in
terms of $H_{p_i}$, $H_{q_iq_j}$, etc.. For a separable Hamiltonian
of the form (\ref{ham}), one can certainly write $T_{i}=H_{p_i}$,
and $V_{ij}=H_{q_iq_j}$, etc., but the latter is not more general
than the former. If the Hamiltonian is not separable, Yoshida's
expression suggests a degree of generality beyond that of the
formalism. It is best to leave the form of the
approximate Hamiltonian function in terms of Poisson brackets,
which is then valid for all Hamiltonians.

For higher order algorithms, the Hamiltonian operator
corresponding to any left-right symmetric factorization is
\bea
\htH &=& \hT+\hV\,
+\,\ep^2\,\left(\,\ct\,[\hT^2\,\hV]+\cv\,[\hV\,\hT\,\hV]\,\right)\,\nn \\
&&+\,\ep^4\,\left(\,\tt\,[\hT\,\hT^3\,\hV]
+ \tv\,[\hV\,\hT^3\,\hV]\right.\, \la{bch4} \\
&&\left. +\,\vt\,[\hT\,(\hT\,\hV)^2]\,
+\,\vv\,[\hV\,(\hT\,\hV)^2]\,\right)\,+\,\dots\, ,
\la{hop4th}
\eea
where $\ct$, $\tv$ etc., are coefficients specific to a particular
algorithm and where we have used the condensed commutator notation
$[\hT^2\hV]\equiv[\hT,[\hT,\hV]]$. Note that for symmetric
decompositions, one has only even order commutators and the
Lie-Poisson correspondence is trivial.
In terms of similarly condensed Poisson brackets,
$\{T^2 V\}\equiv\{T,\{T,V\}\}$, the
Hamiltonian function can be read off by inspection,
\bea
\tH &=& T+V\,
+\,\ep^2\,\left(\,\ct\,\{T^2\,V\}+\cv\,\{V\,T\,V\}\,\right)\,\nn \\
&&+\,\ep^4\,\left(\,\tt\,\{T\,T^3\,V\}
+ \tv\,\{V\,T^3\,V\}\right.\, \nn \\
&&\left. +\,\vt\,\{T\,(T\,V)^2\}\,
+\,\vv\,\{V\,(T\,V)^2\}\,\right)\,+\,\dots\, .
\la{hft4th}
\eea
For the separable
Hamiltonian (\ref{ham}), these higher brackets are:
\bea
\{T\,T^3\,V\}&=&p_ip_jp_kp_lV{ijkl}\, ,\nn\\
\{V\,T^3\,V\}&=&-3p_ip_jV_{ijk}V_k\, ,\nn\\
\{T\,(T\,V)^2\}&=&-2p_i(V_{ikj}V_k+V_{ik}V_{kj})p_j\, ,\nn\\
\{V\,(T\,V)^2\}&=&2 V_iV_{ij}V_j\, .
\la{errsix}
\eea
The results in this section will allow us to analyze any
symplectic algorithm from second to sixth order. Beyond sixth order,
the number of Lie and Poisson brackets proliferates and other
means of determining the Hamiltonian error terms may be more efficient.

\section{Harmonic Oscillator: Second Order Integrator}\la{ho2nd}

To illustrate some of our key ideas in the simplest
context, we will begin our study of the phase error
with the second order factorization scheme
\be
\cT_2(\ep\, ,\,\a)\,\equiv\,
{\rm e}\,^{\fft12\,\ep\,\hT}\,
{\rm e}^{\ep\,\hV_1}\,
{\rm e}\,^{\fft12\,\ep\,\hT}\, ,
\la{tbopt}
\ee
with $\hV_1$ given by
\be
\hV_1=\hV+\a\,\,\ep^2\,[\hV,[\hT,\hV]]\, . \la{vtac2}
\ee
Classically, this Lie commutator produces a modified
force\cite{chin}
\be
[V,[T,V]]=2F_j\fft{\pa F_i}{\pa q_j}\fft{\pa}{\pa p_i}
         =\nabla_i|{\bf F}|^2\fft{\pa}{\pa p_i}\, ,\la{modforce}
\ee
resulting in the following more general second order symplectic integrator
\bea
{\bf q}_1&=&{\bf q}_0+\fft12\,\ep\,\,{\bf p}_0\, ,\nn\\
{\bf p}_1&=&{\bf p}_0+\ep\,\Bigl[{\bf F}({\bf q}_1)+
\a\,\,\ep^2\na|{\bf F}({\bf q}_1)|^2\Bigr]\, , \la{altbsopt} \\
{\bf q}_2&=&{\bf q}_1+\fft12\,\ep\,\,{\bf p}_1\, . \nn
\eea
Here, $({\bf q}_0\,,{\bf p}_0)$ and $({\bf q}_2\,, {\bf p}_1)$ are the
initial and final states of the algorithm respectively. The introduction
of the gradient term with parameter $\a$ will allow us to satisfy
the correctability criterion in its simplest setting. When applied to the
1-D harmonic oscillator with Hamiltonian
\be
H(q,p)=\fft{p^2}2+\fft12\,\w^2\,q^2\, ,\la{h}
\ee
the force gradient is just
\be
F(q)=-\, \w^2\, q\qquad \lra \qquad \na_q|F(q)|^2=2\,\w^4\,q\, .\la{fgra}
\ee

    For the standard Hamiltonian, the approximation Hamiltonian
operator for any symmetric factorization is given by
(\ref{hop4th}).
The non-vanishing error coefficients corresponding to algorithm
(\ref{tbopt}) are just
\be
\ct=-\fft1{24},\quad \cv = \a-\fft1{12}\, ,
\label{tv2}
\ee
\be
\vt=\fft{1}{480}-\fft{1}{24}\a\, ,\quad
\vv=\fft{1}{120}-\fft{1}{6}\a\, .
\la{ctcv}
\ee
The Hamiltonian function is then as given by (\ref{hft4th}).
For the harmonic oscillator as defined by ({\ref{h}),
we have $V_{ij}=\omega^2\delta_{ij}$, $V_{ijk}=0$,
$\{T\,T^3\,V\}=0$, $\{V\,T^3\,V\}=0$ and non-vanishing
brackets,
\bea
\{T,\{T,V\}\}&=&\omega^2 p^2\, , \nn\\
\{V,\{T,V\}\}&=&-\omega^4 q^2\, ,\nn\\
\{T\,(T\,V)^2\}&=&-2\omega^4 p^2\, ,\nn\\
\{V\,(T\,V)^2\}&=&2\omega^6 q^2\, .
\la{harcof}
\eea
Notice the clear separation between the contributions
of the algorithm, which are the error coefficients,
and that of the physical system, which are the Poisson brackets.
The final form of the Hamiltonian
function due to algorithm (\ref{altbsopt}) is therefore,
\bea
\tH(q,p) &=& \fft12\,p^2+\fft12\,\w^2\,q^2
+\,\w^2\,\ep^2\,\left(\,\ct\,p^2-\cv\,\w^2\,q^2\right)\,\nn \\
&&-\, 2\,\w^4\,\ep^4\,\left(\,\vt\,p^2-\,\vv\,\w^2\,q^2\right)
\,+\,\dots\, ,\la{hfct4}\\
&=&\fft1{2\,m^*}\,p^2+\fft12\,k^*\,q^2\, .\la{2ndrha}
\la{hqp3}
\eea
Thus the oscillator being evolved by the algorithm is one with
an effective mass and spring constant,
\bea
m^*=m^*(\ep)&\equiv&(1+2\,\ep^2\,\w^2\,\ct
-4\ep^4\w^4\vt+\dots\,)^{-1}
     \, ,\la{calp}\\
k^*=\,\,k^*(\ep)&\equiv&\,(1-2\,\ep^2\,\w^2\,\cv
+4\ep^4\w^4\vv\,+\dots\,)\,\w^2
     \, ,\la{calq}
\eea
from which one can deduce the approximate angular frequency
\be
\tw(\ep)=\sqrt{\fft{k^*}{m^*}}\, .\la{apw}
\ee
The phase error is simply related to the fractional deviation of the the
approximate angular frequency from the exact frequency:
\be
\D\f=(\w_A-\w){\rm T}=2\pi({{\w_A}\over{\w}} -1)\, .\la{phase}
\ee
This is the fundamental thrust of our analysis: tracking the phase error of
the algorithm back to its factorization coefficients.
Observe now that from (\ref{calp}) and (\ref{calq}), we have
\bea
\tw(\ep) &=&\w\sqrt{(1+2\,\ep^2\,\w^2\,\ct+\,\dots\,)
  (1-2\,\ep^2\,\w^2\,\cv+\,\dots\,)}\, ,\\
&=&\w\Bigl[1+\ep^2\,\w^2(\ct-\cv)+O(\ep^4)\Bigr]\, .\la{wappro}
\eea
In general, the approximate frequency is second order in error, as
befitting a second order algorithm. However, if the correctability
criterion $\ct=\cv$ is satisfied, then $\tw$ is fourth order.
Moreover, if the algorithm is originally fourth order with
$\ct=\cv=0$ then satisfying $\vt=\vv$ would make $\tw$ sixth
order. Thus an $n$th algorithm can have an $(n+2)$th order phase
error if its error coefficient satisfies the correctability
criterion. This is the key connection linking the phase error with
correctable algorithms. (Note that by making $\ct=\cv$ (but not
zero) and $\vt=\vv$, would not make the phase error sixth order.)

With only one free parameter presently
available, we can only set $\ct=\cv=-{1\over{24}}$ with the choice
\be
\a=\fft1{24}\, ,\la{alim}
\ee
thus making $\tw$ fourth order. This particular value corresponds to
the well known propagator first
derived by Takahashi and Imada\cite{ti} for computing the quantum
statistical trace\cite{ti} to fourth order. The same factorization scheme,
interpreted as symplectic corrector algorithm (\ref{altbsopt}), has also
been used by Lopez-Marcos {\it et al.}\cite{mar96,mar97} and
Wisdom {\it et al.}\cite{wis96} for solving classical and celestial
dynamical problems. With this choice of $\a$, the coefficient of the
fourth order frequency error is, from (\ref{calp}), (\ref{calq})
and (\ref{ctcv}),
\bea
\fft{\w^{(4)}}{\w}&=&
\lim_{\ep\ra\, 0}\,\left[\fft1{\ep^4}
\left(\fft{\tw}{\w}-1\right)\right]\, ,\nn\\
&=&2\w^4(\vv-\ct^2-\vt)
=-\,\fft{\w^4}{720}\, .\la{prefran2nd4}
\eea

To gauge the relative importance of this phase error, let's compare it
to the energy error after one period.
Since it is the modified, or approximate Hamiltonian that is conserved by
the algorithm, \ie
\be
\tH(q,p)=\tH(q_0,p_0)\, ,\la{apprcons}
\ee
the energy after one period $\mathrm{T}=2\pi/\w$ can be expressed as
\be
H(q_{_{\mathrm{T}}},p_{_{\mathrm{T}}})
=H(q_0,p_0)+\ep^2\,\D H^{(2)}_{\,\mathrm{T}}(\ep^2)
+\ep^4\,\D H^{(4)}_{\,\mathrm{T}}(\ep^2)+\ep^6\,
\D H^{(6)}_{\,\mathrm{T}}(\ep^2)+O(\ep^8)\, . \la{2ndenerr}
\ee
From (\ref{hfct4}), we have in particular,
\bea
\D H^{(2)}_{\,\mathrm{T}}(\ep^2)
&=&-\hs\left.\w^2\,\left(\ct\,(p^2-p_0^2)-\,\cv\,\w^2\,
(q^2-q_0^2)\right)\,\right|_{\,\,t=\mathrm{T}}
\, ,\la{2ndenerfrasol}\\
\D H^{(4)}_{\,\mathrm{T}}(\ep^2)
&=&\left.2\,\w^4\,\left(\vt\,(p^2-p_0^2)-\,\vv\,\w^2\,
(q^2-q_0^2)\right)\,\right|_{\,\,t=\mathrm{T}}\, .\la{4theng}
\eea
In order to compute these energy deviation errors, we must solve for
$p(t)$ and $q(t)$ according to Hamiltonian $H_A$:
\be
\pmatrix{q(t;\ep) \cr p(t;\ep) }=
\pmatrix{\cos(\w_A t) & (m^*\w_A)^{-1}\sin(\w_A t)\cr
          -(m^*\w_A)\sin(\w_A t)& \cos(\w_A t) }
\pmatrix{q_0 \cr p_0}\, . 
\la{solut}
\ee
Since $m^*$ and $\w_A$ are $\ep^2$-dependent, each function
$\D H^{(n)}(\ep^2)$ contains further dependence on $\ep^2$. We now
define the constant energy error coefficients $E^{(n)}_T$ via
\be
H(q_{_{\mathrm{T}}},p_{_{\mathrm{T}}})
-H(q_0,p_0)\equiv\D E_{\,\mathrm{T}}=\ep^2\,E^{(2)}_{\,\mathrm{T}}
+\ep^4\,E^{(4)}_{\,\mathrm{T}}+\ep^6\,E^{(6)}_{\,\mathrm{T}}
+O(\ep^8)\, , \la{engerr}
\ee
where for example, we have
\bea
E^{(2)}_{\,\mathrm{T}} &=& \D H^{(2)}_{\,\mathrm{T}}(0)\, ,\nn\\
E^{(4)}_{\,\mathrm{T}} &=& \D H^{(4)}_{\,\mathrm{T}}(0)
+\D H^{(2)\prime}_{\,\mathrm{T}}(0)\, ,\nn\\
E^{(6)}_{\,\mathrm{T}} &=& \D H^{(6)}_{\,\mathrm{T}}(0)
+\D H^{(4)\prime}_{\,\mathrm{T}}(0)
+{1\over {2!}}\D H^{(2)\prime\prime}_{\,\mathrm{T}}(0)\, ,\nn\\
E^{(8)}_{\,\mathrm{T}} &=& \D H^{(8)}_{\,\mathrm{T}}(0)
+\D H^{(6)\prime}_{\,\mathrm{T}}(0)
+{1\over {2!}}\D H^{(4)\prime\prime}_{\,\mathrm{T}}(0)
+{1\over {3!}}\D H^{(2)\prime\prime\prime}_{\,\mathrm{T}}(0)\, .
\la{eng28}
\eea
Here, {\it the prime denotes derivative with respect to $\ep^2$}.
From the form of each $\D H^{(n)}_{\,\mathrm{T}}(\ep^2)$,
since $\ep=0$ implies that
$\w_A=\w$, $p(\mathrm{T})=p_0$ and $q(\mathrm{T})=q_0$,
we must have
\be
\D H^{(n)}_{\,\mathrm{T}}(0)=0\, ,\la{dh}
\ee
and therefore
\be
E^{(2)}_{\,\mathrm{T}}=0\, .
\la{2ndenerfrasol2}
\ee
Thus for periodic motion, despite the fact the algorithm is only second
order, the energy error is actually fourth order after one
period.

The fourth order energy error is given by
\bea
E^{(4)}_T &=& \D H^{(2)\prime}_{\,\mathrm{T}}(0)
=-2\,\w^2\,(\ct p_{\mathrm{T}}\,p_{\mathrm{T}}\,^\prime
-\cv\w^2 q_{\mathrm{T}}\,q^\prime_{\mathrm{T}})\Bigr|_{\,\ep=0}\,\, ,\nn\\
&=&4\pi\w^5p_0\,q_0\,(\ct-\cv)(\ct+\cv)\, ,
\la{eng4cof}
\eea
where we have used
\be
q^\prime({\mathrm{T}};0)=\frac1{\w}\,p_0\,\w_A^\prime(0){\mathrm{T}}
\quad{\rm and}\quad
p^\prime({\mathrm{T}};0)=-\w\,q_0\,\w_A^\prime(0){\mathrm{T}}\, ,\nn
\ee
and from (\ref{wappro}),
\be
\w_A^\prime(0){\mathrm{T}}=2\pi \w^2(\ct-\cv)\, .
\la{waprime}
\ee
The fourth order error now vanishes if the algorithm satisfies
the correctability criterion $\ct=\cv$.
Thus for a {\it correctable} second order algorithms,
after each period, the phase error is
fourth order and the energy error is sixth order.

Since the factor (\ref{waprime}) is common to all first derivatives
(in $\ep^2$), we conclude that for $\ct=\cv$
\be
\D H^{(n)\prime}_{\,\mathrm{T}}(0)=0\, .
\la{dhp}
\ee
Hence for $\ct=\cv$, the sixth order energy error can be now computed as
\bea
E^{(6)}_{\mathrm{T}}&=&
\fft12\,\D H^{(2)\prime\prime}_{\,\mathrm{T}}(0)\, ,\la{e6}\\
&=& 2\pi\w^6\Bigl[2\pi(p_0^2-\w^2q_0^2)-p_0\,q_0\w\Bigr](\ct+\cv)(\ct-\cv)^2\nn\\
&&-4\pi p_0\,q_0\w^7(\ct+\cv)\Bigl[2(\vt-\vv)+\ct^2+\cv^2\Bigr]\, ,\nn\\
&=&{{\pi\w^7}\over{2160}}p_0\,q_0\, . \la{e62}
\eea

The above calculation demonstrates the general property of the
energy deviation error after one period.
For correctable algorithms, the first two terms in
the error expansion (\ref{eng28}) vanish identically, which means that
to compute $E^{(6)}_{\mathrm{T}}$, one need not know the explicit
form $\D H^{(6)}_{\,\mathrm{T}}(\ep^2)$.
However, in order to compute $\D H^{(2)\prime\prime}_{\,\mathrm{T}}(0)$,
one must know $m^*(\ep^2)$ and $\w_A(\ep^2)$ accurately to $O(\ep^4)$,
which means knowing the fourth order Hamiltonian error function,
or $\D H^{(4)}_{\,\mathrm{T}}(\ep^2)$. Thus although (\ref{e6}) makes
no reference to $\D H^{(4)}_{\,\mathrm{T}}(\ep^2)$, one must know it
implicitly.
Similarly, $E^{(8)}_{\mathrm{T}}$ can be computed
from $\D H^{(2)}_{\,\mathrm{T}}(\ep^2)$
and
$\D H^{(4)}_{\,\mathrm{T}}(\ep^2)$ via
\be
E^{(8)}_{\mathrm{T}}=
{1\over {2!}}\D H^{(4)\prime\prime}_{\,\mathrm{T}}(0)
+{1\over {3!}}\D H^{(2)\prime\prime\prime}_{\,\mathrm{T}}(0)\, .
\la{e8}
\ee
However, in order to compute $\D H^{(2)\prime\prime\prime}_{\,\mathrm{T}}(0)$
one must know $\D H^{(2)}(\ep^2)$ correctly to $O(\ep^6)$. This would again
require knowing the sixth order error Hamiltonian or $\D H^{(6)}(\ep^2)$.
In general, $E^{(n)}_{\mathrm{T}}$ can be compute two orders beyond the
accuracy of knowing the Hamiltonian.

    To summarize, for a second order algorithm, the energy after one period 
is automatically fourth order in $\ep$ ($=\Delta t$). If the algorithm
is correctable, then the energy error is sixth order. For special initial
conditions $p_0=0$ or $q_0=0$, by solving the
algorithm exactly in the case of the harmonic oscillator\cite{chinscuro}, 
one can show that the energy error is 
actually tenth order. This last error reduction only occurs
for the harmonic oscillator. Nevertheless 
this further emphasizes that the energy error after one period is not a very 
good gauge of any integrator's accuracy. On the other hand, the phase error, 
as reflected in the fractional change of the oscillator's angular 
frequency, can at most be fourth order and is a much more stringent 
and discriminating benchmark.

\section{Harmonic Oscillator: Fourth Order Forward Integrators}\la{ho4th}
   
   Beyond second order, all symplectic algorithms of the form (\ref{prod})
must have some negative intermediate time
steps\cite{sheng,suzukinogo,goldman}.
This means that at some intermediate time, the algorithm is moving
the phase trajectory backward in time. For classical mechanics, which is
time-reversible, these negative time steps are harmless. However for
solving time-irreversible problems, such as the diffusion or Fokker-Planck
equation, backward time step evolution is not possible. These systems can
only be solved by {\it forward} decomposition algorithms, with all
positive, even intermediary, time steps. Some fourth
order forward algorithms have been derived recently for solving a variety of
time-irreversible\cite{lang4,dmc4}, and
time-reversible\cite{chin,chinchen02,chinchen03}
equations, both with excellent results. Beyond second order,
purely forward time steps are possible only if one
include the commutator $[\hV,[\hT,\hT]]$ in addition to operators
$\hT$ and $\hV$ in the factorization process. In this work we will apply
these fourth order forward algorithms to study the phase problem of
periodic motion. In this section, we further generalize our study of
the harmonic oscillator by use of these fourth order forward algorithms.

   Chin and Chen\cite{chinchen02,chinchen03} have introduced a family
of fourth order forward algorithms $4ACB$ parametrized by a parameter $t_0$.
We use here a slightly generalized form by multiplying the central commutator
by $1-\a$ and adding $\a/2$ times the commutator to each
potential operator on each side. The resulting algorithm has the operator
form
\be
{\cal T}_{ACB}^{(4)}(\ep\, ,\,\a)\equiv
  {\rm e}^{ t_0\,\ep\, \hT}
  {\rm e}^{ v_1\,\ep\, \hV_1}
  {\rm e}^{ t_1\,\ep\, \hT}
  {\rm e}^{ v_2\,\ep\, \hV_2}
  {\rm e}^{ t_1\,\ep\, \hT}
  {\rm e}^{ v_1\,\ep\, \hV_1}
  {\rm e}^{ t_0\,\ep\, \hT}\, , \la{algacop}
\ee
where
\bea
\hV_1&=&\hV+\fft{\a}2\,\fft{u_0}{v_1}\,\ep^2\,[\hV,[\hT,\hV]]\, , \nn\\
\hV_2&=&\hV+(1-\a)\,\fft{u_0}{v_2}\,\ep^2\,[\hV,[\hT,\hV]]\, , \la{vtac3}\\
u_0&=&{1\over 12}\biggl[1-{1\over{1-2t_0}}+{1\over{6(1-2t_0)^3}}\biggr]\, ,
\eea
and
\be
t_1={1\over 2}-t_0\, ,\quad
v_1={1\over 6}{1\over{(1-2 t_0)^2}}\, ,\quad
v_2=1-2v_1\, .\la{acofac}
\ee
The corresponding forward symplectic integrator can be read
off directly as
\bea
{\bf q}_1&=&{\bf q}_0+\ep\,t_0\,{\bf p}_0\, ,\nn\\
{\bf p}_1&=&{\bf p}_0+\ep\,\Bigl[v_1\,{\bf F}({\bf q}_1)+
\fft{\a}2\,u_0\,\ep^2\na|{\bf F}({\bf q}_1)|^2\Bigr]\, ,\nn\\
{\bf q}_2&=&{\bf q}_1+\ep\,t_1\,{\bf p}_1\, ,\nn\\
{\bf p}_2&=&{\bf p}_1+\ep\,\Bigl[v_2\,{\bf F}({\bf q}_2)+
(1-\a)\,u_0\,\ep^2\na|{\bf F}({\bf q}_2)|^2\Bigr]\, ,\la{facb} \\
{\bf q}_3&=&{\bf q}_2+\ep\,t_1\,{\bf p}_2\, ,\nn\\
{\bf p}_3&=&{\bf p}_2+\ep\,\Bigl[v_1\,{\bf F}({\bf q}_3)+
\fft{\a}2\,u_0\,\ep^2\na|{\bf F}({\bf q}_3)|^2\Bigr]\, ,\nn\\
{\bf q}_4&=&{\bf q}_3+\ep\,t_0\,{\bf p}_3\, ,
\nn
\eea
where $({\bf q}_0\,,{\bf p}_0)$ and $({\bf q}_4\,, {\bf p}_3)$ are the
initial and final states of the algorithm respectively.
The parameter $\a$ can be changed from 0 to 1, but there is really
no restriction on its range. When applied to the harmonic
oscillator, the parameter $\alpha$ can be used to correct the
algorithm to sixth order. The parameter $t_0$ can be varied from
0 to $t_c={1\over 2}(1-{1\over{\sqrt 3}})\approx 0.21$.
For $t_0=0$, the final force evaluation can be reused at the next
iteration, thus eliminating one force evaluation.
At the upper limit of $t_0=t_c$, $v_2=0$, also eliminates one
force evaluation. For $t_0>t_c$, $v_2$ becomes negative, and the
algorithm ceases to be a forward algorithm.

    Our analysis of the second order algorithm can now be repeated
verbatim for the fourth order case. The approximate Hamiltonian
operator corresponding to any symmetric fourth order algorithm is of the
form,
\bea
\htH= \hT+\hV +\ep^4 &\Bigl( & \tt [\hT\hT^3\hV]
+ \tv[\hV\hT^3\hV] \la{bch4g} \\
&&+\vt[\hT(\hT\hV)^2]
+\vv[\hV(\hT\hV)^2]\,\,\,\Bigr)+O(\ep^6)\, . \nn
\eea
For the harmonic oscillator, $[\hT^3\hV]=0$, and the first two
error term vanishes identically. The evaluation of the last two
error coefficients for the family of fourth order algorithm
(\ref{facb}) is non-trivial and is given Appendix \ref{4th1}.
The corresponding Hamiltonian function, after recalling the Poisson
form (\ref{hft4th}) and brackets (\ref{harcof}), is
\bea
H_A(q,p)
&=&\fft{p^2}2+\fft12\,\w^2\,q^2-2\,\w^4\,\ep^4\,
\left(\,\vt\,p^2-\vv\,\w^2\,q^2\right)+\,\dots\, ,\la{4}\\
&=& \fft1{2\,m^*}p^2+\fft12 k^*q^2\, ,\la{4thrha}
\la{4thhamop}
\eea
with
\bea
m^*=m^*(\ep)&\equiv&(1-4\,\ep^4\,\w^4\,\vt+\,\dots\,)^{-1}
     \,\,\, ,\la{m4}\\
k^*=\,\,k^*(\ep)&\equiv&\w^2\,(1+4\,\ep^4\,\w^4\,\vv+\,\dots\,)
     \, ,\la{k4}
\eea
and approximate frequency
\bea
\tw(\ep)
&=&\w\sqrt{(1+4\,\ep^4\,\w^4\,\vv+\,\dots\,)
  (1-4\,\ep^4\,\w^4\,\vt+\,\dots\,)}\, ,\\
&=&\w\Bigl[1+2\ep^4\,\w^4(\vv-\vt)+O(\ep^6) \Bigr]\, .
\la{wappro4}
\eea
Again, one immediately sees that if the sixth order correctability criterion
\be
\vv=\vt\, ,
\ee
is satisfied, then $\tw$ will be sixth order.
Note that now we have
\bea
\w_A^\prime {\mathrm{T}}\Bigr|_{\,\ep=0}&=&0\, ,\nn\\
\w_A^{\prime\prime} {\mathrm{T}}\Bigr|_{\,\ep=0}&=&4\pi \w^4(\vt-\vv)\, ,
\la{wa4pp}
\eea
where primes still denote derivative with respect to $\ep^2$.
The conservation of $\tH(q,p)$ again implies that the energy deviation
after one period can be expressed as
\be
H(q_{_{\mathrm{T}}},p_{_{\mathrm{T}}})
=H(q_0,p_0)
+\ep^4\,\D H^{(4)}_{\,\mathrm{T}}(\ep^2)
+\ep^6\,\D H^{(6)}_{\,\mathrm{T}}(\ep^2)
+\ep^8\,\D H^{(8)}_{\,\mathrm{T}}(\ep^2)
+O(\ep^{10})\, , \la{4thenerr}
\ee
with
\be
\D H^{(4)}_{\,\mathrm{T}}(\ep^2)
=\left.2\,\w^4\,\left(\vt\,(p^2-p_0^2)-\,\vv\,\w^2\,
(q^2-q_0^2)\right)\,\right|_{\,\,t=\mathrm{T}}\, .
\la{ffpeerr}
\ee
The constant energy error coefficients $E^{(n)}_T$ defined by
\be
H(q_{_{\mathrm{T}}},p_{_{\mathrm{T}}})
-H(q_0,p_0)\equiv\D E_{\,\mathrm{T}}
=\ep^4\,E^{(4)}_{\,\mathrm{T}}
+\ep^6\,E^{(6)}_{\,\mathrm{T}}
+\ep^8\,E^{(8)}_{\,\mathrm{T}}
+\ep^{10}\,E^{(10)}_{\,\mathrm{T}}
+O(\ep^{12})\, , \la{eng4err}
\ee
are now of the form
\bea
E^{(4)}_T &=&
\D H^{(4)}_{\,\mathrm{T}}(0)\, ,\nn\\
E^{(6)}_T &=&
\D H^{(6)}_{\,\mathrm{T}}(0)
+\D H^{(4)\prime}_{\,\mathrm{T}}(0)\, ,\nn\\
E^{(8)}_T &=&
\D H^{(8)}_{\,\mathrm{T}}(0)
+\D H^{(6)\prime}_{\,\mathrm{T}}(0)
+{1\over {2!}}\D H^{(4)\prime\prime}_{\,\mathrm{T}}(0)\, ,\nn\\
E^{10}_T &=&
\D H^{(10)}_{\,\mathrm{T}}(0)
+\D H^{(8)\prime}_{\,\mathrm{T}}(0)
+{1\over {2!}}\D H^{(6)\prime\prime}_{\,\mathrm{T}}(0)
+{1\over {3!}}\D H^{(4)\prime\prime\prime}_{\,\mathrm{T}}(0)\, .
\la{eng48}
\eea
Now, because of (\ref{wa4pp}), for $\vt=\vv$, not only we do have
$\D H^{(n)}_{\,\mathrm{T}}(0)=0$, but also
\be
\D H^{(n)\prime}_{\,\mathrm{T}}(0)=0\quad
{\rm and}\quad
\D H^{(n)\prime\prime}_{\,\mathrm{T}}(0)=0\, .
\la{pandpp}
\ee
This implies that
\be
E^{(4)}_T=E^{(6)}_T=E^{(8)}_T=0\, ,
\ee
and the first non-vanishing energy error is tenth order,
\be
E^{10}_T =
{1\over {3!}}\D H^{(4)\prime\prime\prime}_{\,\mathrm{T}}(0)\, .
\ee
However, as noted in the last section, in order to compute this,
one must determine the sixth order error Hamiltonian.

Due the complexicity of the algorithm, these higher error terms
are difficult to compute by Lie series. However, they
can always be computed using the matrix method\cite{chinscuro}.
For brevity, we will skip over the details and just report 
the final results.

We have shown earlier that the fourth order phase error term will 
vanish if $\vt=\vv$. For a given value of $t_0$, this criterion 
can now be satisfied by a specific choice of $\a$ given by $\a=\a(t_0)$ in
(\ref{alpha}). Using this functional form to eliminate $\a$ in terms of
$t_0$, the sixth order error term $\fft{\w^{(6)}}{\w}=f(t_0)$
scaled such that $\w=1$, is plotted in Fig.\ref{fig6th}.

\begin{figure}[H]
\centering
\epsfxsize=4.5in
\hspace*{0in}
\epsffile{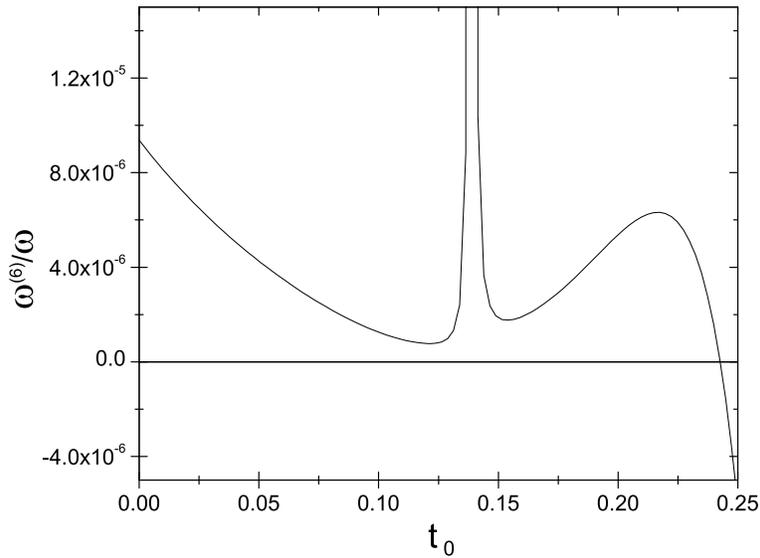}
\caption{The sixth order angular frequency error as a function of the algorithm's
parameter $t_0$.}
\la{fig6th}
\end{figure}

Within the forward range of $0\leq t_0 \leq 0.21$, the sixth order
frequency error has a minimum of value
\be
\left.
\fft{\w^{(6)}}{\w}\right|_{\,\,min}=\,7.718621317057857\times10^{-7}\,\w^6\, ,
\la{6therrmin}
\ee
at $t_0=0.12129085056575276$, and a pole at $t_0=0.13882413776781183$.
Note that outside of the forward range, the error can actually vanish
at $t_0=0.24265927253055103$.

    The eighth order energy deviation error after one period is
\be
\D E^{(8)}_{\,\mathrm{T}}=16\,\pi\,\w^9\,(\vt^2-\vv^2)
\,q_0\,p_0\, ,\la{4thhvan1}\\
\ee
which again vanishes for $\,\,\vt=\vv\,\,$ {\it or} $\,\,\vt=-\,\vv$,
analogous to the second order case.

   Thus for a corrected fourth order algorithm, the first non-zero
energy deviation error is tenth order. This is plotted in
Fig.{\ref{fig6thh} scaled such that $\w=q_0=p_0=1$.

\begin{figure}[H]
\centering
\epsfxsize=4.5in
\hspace*{0in}
\epsffile{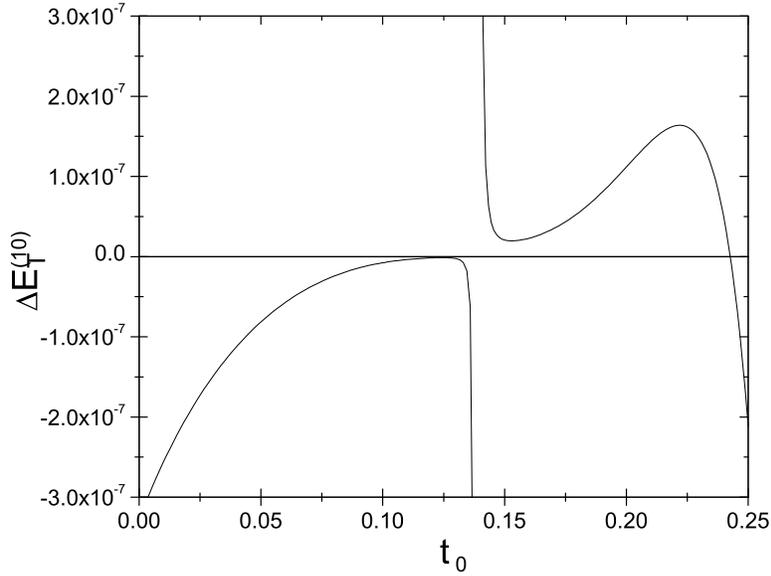}
\caption{The tenth order energy deviation error after one period
as a function of the algorithm's parameter $t_0$.}
\la{fig6thh}
\end{figure}

   Within the forward range of $0\leq t_0 \leq 0.21$, the tenth order
energy deviation error has a minimum of value
\be
\left.
\D E^{(10)}_{\,\mathrm{T}}\right|_{\,\,min}
=-1.3398713813012635\times10^{-9}\,\w^{11}\,q_0\,p_0\, , \la{4thherrmin}
\ee
at $t_0=0.12482248354859667$, and a pole at $t_0=0.13882413776781183$ 
(same as in the frequency case). In both cases the error term vanishes 
at the same value, \ie $t_0=0.24265927253055103$, outside of
the forward range. (Note also that this error term vanishes for
special starting value of $p_0=0$ or $q_0=0$. It can be shown
that for either $p_0=0$ or $q_0=0$, the first non-vanishing
energy error term is $16^{th}$ order, again demonstrating that
the phase error dominates overwhelmingly over the energy error.)

\section{The 2-D Kepler Problem}\la{kepl}

In light of our previous discussion, for long term trajectory simulation,
one must judge all symplectic algorithms on how well they minimize
the phase errors rather than the energy error. In this section, we will
examine Keplerian motions in 2-D defined
by the Hamiltonian
\be
H={1\over 2}{\bf p}^2-{1\over{|\bf q}|}\, .
\label{kepler}
\ee
Here, our analysis of fourth order algorithms  will not be
as extensive as in the  harmonic oscillator case because the
approximate Hamiltonian
\bea
\htH= \hT+\hV +\ep^4&\Bigl(&\tt[\hT\hT^3\hV]
\,+\, \tv[\hV\hT^3\hV] \la{bch4h} \\
&&+\,\,\vt[\hT(\hT\hV)^2]
\,+\,\vv[\hV(\hT\hV)^2]\,\,\,\Bigr)+O(\ep^6)\, , \nn
\eea
can no longer be solved analytically.
The operator
$[\hT^3\hV]\neq 0$ and while we can still force
$\vt=\vv$ as in the harmonic oscillator case, we have no way of
ensuring that $\tt=\tv$. Currently, there are no known
fourth order forward symplectic algorithms that can be corrected
to sixth order. Nevertheless, identical analysis as
in the harmonic oscillator case shows that
\be
E^{(4)}_T =
\D H^{(4)}_{\,\mathrm{T}}(0)=0\, ,
\la{eng4k}
\ee
and the energy error after one period must be at least
sixth order. Thus if fourth order algorithms are used to
solve Keplerian orbits, it is more fitting to examine their
fourth order phase errors instead.

For two-dimensional motion, there are two basic phase angles
associated with the two sets of canonical variables $(q_1,p_1)$
and $(q_2,p_2)$. A convenient measure of these phase errors is
the precession error of the orbit in the $(q_1,q_2)$ plane, which can
be tracked\cite{hiord} by the rotation of the Laplace-Runge-Lenz
(LRL) vector
\be
{\bf A}={\bf p}\times{\bf L}-\hat{\bf q}\, .
\ee
In the above definition, ${\bf L}={\bf q}\times{\bf p}$,
is the angular momentum vector.

To see how various algorithms compare, we first plot the fourth order
energy error function defined by
\be
H_4({\bf q}(t),{\bf p}(t)) =
\lim_{\ep\rightarrow 0}{1\over{\ep^4 E_0}}[E({\bf q}(t),{\bf p}(t))-E_0]\, ,
\ee
in Fig.\ref{figene}. 

\begin{figure}[H]
\centering
\epsfxsize=4.5in
\hspace*{0in}
\epsffile{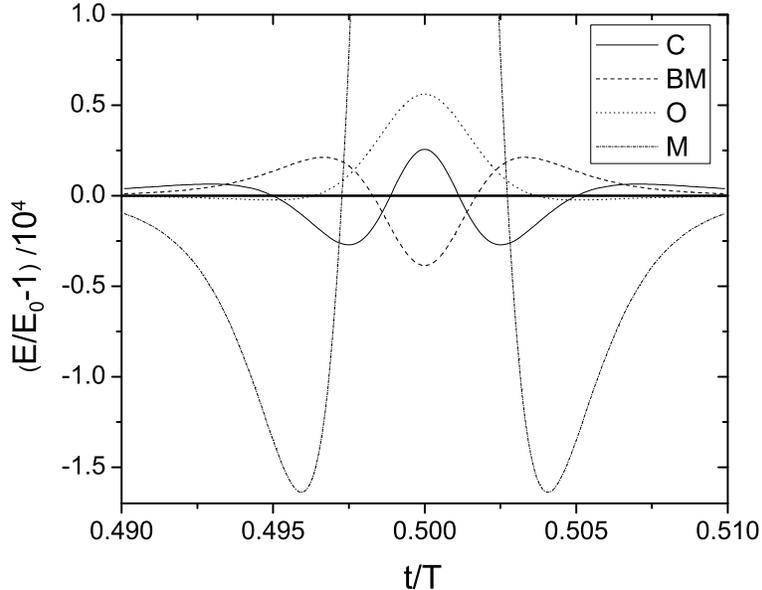}
\caption{The energy error at half a period for an eccentricity of 0.9}
\la{figene}
\end{figure}

Note that this is an intrinsic function characteristic of each algorithm
independent of the step size. We compute this function by finding the 
energy deviation from the initial energy along the orbit and then dividing 
it by $\ep^4$. As $\ep$ gets smaller and smaller, this function converges 
to its limiting form. The functional form is basically unchanged
for $\ep\le {\mathrm T}/3000$, where ${\mathrm T}$ is the period of the 
Keplerian orbit. All results shown in Fig.\ref{figene} are computed with 
$\ep={\mathrm T}/5000$.

Since we have shown that $E({\bf q}({\mathrm
T}),{\bf p}({\mathrm T}))-E_0=O(\ep^6)$, $H_4$ vanishes 
exactly after one period. Thus each of energy error curve of Fig.\ref{figene}
reverts back to zero at $t={\mathrm T}$. This is a characteristic 
behavior of all symplectic algorithms. Non-symplectic Runge-Kutta 
algorithms do not have this property and their energy deviation error 
accumulates rather than vanishing after each period. However, 
even for symplectic algorithms, the energy deviation error
is non-vanishing at other times.
Here, due to the high eccentricity ($e=0.9$) of the orbit, the
energy error is at a maximum near mid-period. Algorithm Chin-C
(C), is the forward algorithm (\ref{algacop}) with $t_0=1/6$ and
$\alpha=0$, first derived in {\cite{chin}}; Blanes-Moan
(BM) is an algorithm recommended in McLachlan and Quispel's
review\cite{mcl02}; Omelyan {\it et al.}\cite{ome03}(O) is a
recent alternative forward algorithm that uses the same force
gradient defined by (\ref{fgra}); McLachlan\cite{mcl95}(M) is a 
greatly improved version of the first fourth order 
Ruth-Forest\cite{forest} algorithm. With the exception of M, all
algorithms have comparable error height at mid-period. Note
however that BM requires six force evaluations, M uses four force
evaluations, O uses four force plus four force-gradient
evaluations, but C uses only three force and one force-gradient
evaluation. Algorithm M's error height reaches up to 14, which is
more than twenty times higher. This is rather surprising, since
algorithm M works very well in solving quantum
mechanical\cite{chinchen02,serna} and three-body\cite{chinchen03}
problems.

    In Fig.\ref{figlrl2}, we track the rotation of the LRL vector 
during orbital motion.

\begin{figure}[H]
\centering
\epsfxsize=4.5in
\hspace*{0in}
\epsffile{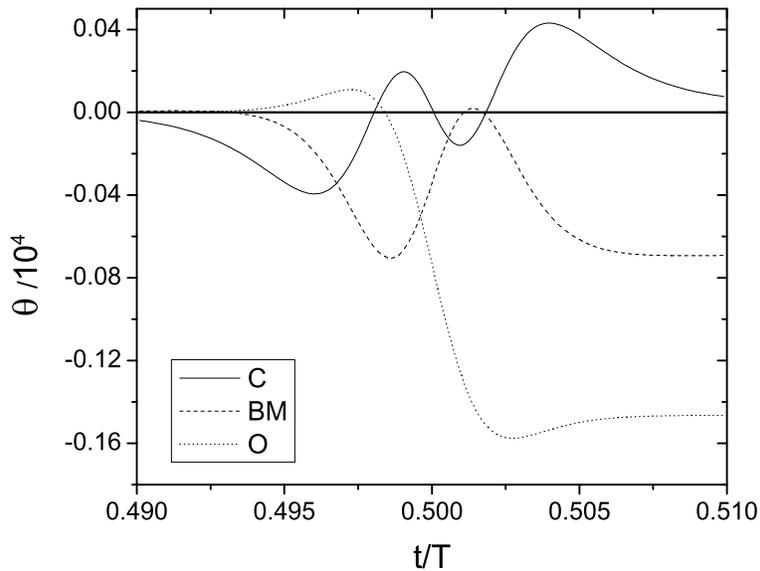}
\caption{The precession deviation error after half a period for eccentricity 0.9 
with starting point $q=(10,0)$ and $p=(0,0.1)$}
\la{figlrl2}
\end{figure}

If the orbit is exact, the LRL vector is a constant vector
pointing along the semi-major axis of the orbit. If the orbit
precesses, then the LRL vector rotates accordingly.
At any point in the orbit, the angle of the LRL vector is
given by
\begin{equation}
\theta(t)=\tan^{-1}\biggl[{{A_y(t)}\over{A_x(t)}}\biggr]=
\epsilon^4\theta_4(t)+\ep^6\theta_6(t)+\cdots\, ,
\end{equation}
and from which one can extract the fourth order angle error
function via
\begin{equation}
\theta_4(t)=\lim_{\epsilon\rightarrow 0}
{1\over {\epsilon^4}}\theta(t)\, .
\end{equation}
Again, this intrinsic function is computed in the limit of
small $\ep$. We have checked that it has indeed converged to
its limiting form for $\ep={\mathrm T}/5000$.
Since the orbit precesses the most when the particle
is closest to the attractor, the LRL vector rotates measurably
only during mid-period. It is constant before, and remained constant
after the mid-period. Thus the rotation after one period is essentially 
the same as the rotation shortly after mid-period. Note that this 
(phase) angle error {\it do not} revert back to zero after each period, 
but accumulate after each period even for symplectic algorithms regardless 
of order. Thus the only way to minimize this phase error is to make it as 
small as possible. From Fig.\ref{figlrl2}, we see that algorithm 
C's rotation angle after mid-period in nearly an order of magnitude 
smaller that that of either BM or O. The actual values after one 
period are: 0.0076, -0.0692, -0.1466 respectively. Algorithm M's rotation 
function reaches down to $\approx -2.5$, which is an order of magnitude 
greater than that of BM and O and two orders of magnitude greater than 
that of C. We did not bother to plot it.

Since parameters $t_0$ and $\alpha$ are at our disposal, we can
further optimize the family of algorithm (\ref{algacop}) to reduce
the rotation error. The resulting optimal choice is shown in
Fig.\ref{figcop2}, with $t_0=0.166160$ and $\alpha=0$. The angle
error after one period is further reduced by a factor of five from
$0.0360$ to $0.0077$.

\begin{figure}[H]
\centering
\epsfxsize=4.5in
\hspace*{0in}
\epsffile{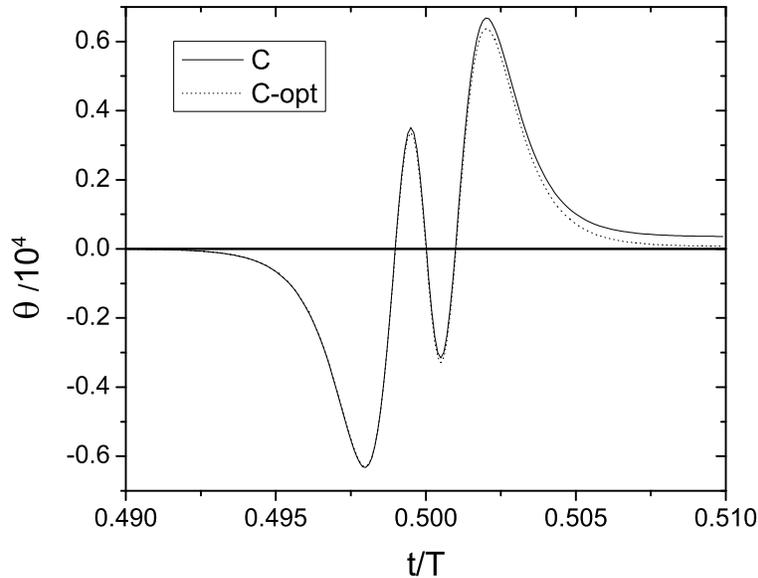}
\caption{The precession deviation error after half a period for eccentricity 0.936 
with starting point $q=(10,0)$ and $p=(0,0.08)$}
\la{figcop2}
\end{figure}

While one can optimize the family of algorithm (\ref{algacop}) for
any one specific problem, or at one eccentricity, it is of greater
value to devise an optimal algorithm for solving a general class
of problems. For the Kepler problem, all possible shapes of closed
orbits are spanned by the eccentricity; it is thus more desirable
if one can devise an optimal algorithm for all values of the
eccentricity. In Fig.\ref{figecc}, we plot the LRL rotation angle
after one period as a function of the orbit's eccentricity, as
determined by different initial conditions. 

\begin{figure}[H]
\centering
\epsfxsize=4.5in
\hspace*{0in}
\epsffile{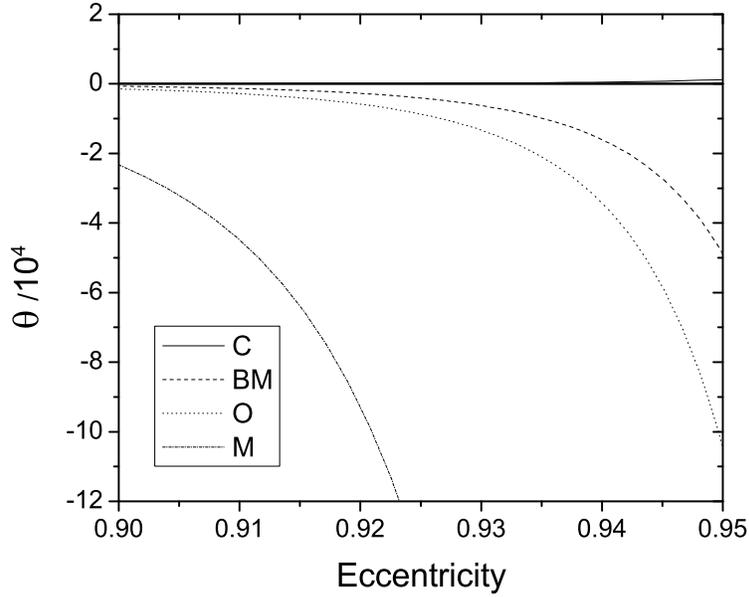}
\caption{The precession deviation error for highly eccentric orbits.}
\la{figecc}
\end{figure}

Most algorithms work well for orbits of low eccentricity and the rotation 
angle is correspondingly small. We therefore compare algorithm at $e\ge
0.9\,$. At $e=0.95$, the angle error values for M, BM, O and C are
respectively -166.1870, -4.8865, -10.4470 , and 0.1244. Algorithm
C's angle error is orders of magnitude smaller than other algorithms.

   In Fig.\ref{figeccopt}, we again show that a better algorithm can
be devised from the family of algorithms (\ref{algacop}). The
choice of $\alpha=0$ (only one force-gradient), and $t_0=0.166160$
(only slightly below the canonical value of $t_0=1/6$), produces
an algorithm with uniformly small phase error up to $e=0.95\,$. At
$e=0.95$ the angle error value for Opt-C is -0.00357, compares to
C's value of 0.12363.

\begin{figure}[H]
\centering
\epsfxsize=4.5in
\hspace*{0in}
\epsffile{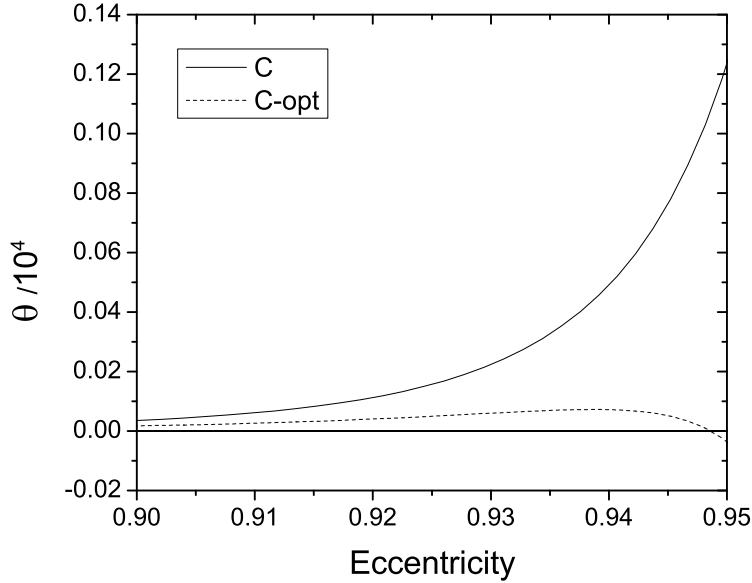}
\caption{The precession deviation error for highly eccentric orbits.}
\la{figeccopt}
\end{figure}

\section{Conclusion}\la{conclusion}

    In this work we showed that for periodic motion, the energy error after 
one period is generally two orders higher than that of the algorithm.
If the algorithm is correctable, the phase error can also be reduced two 
orders higher. The use of fourth order forward time step integrators can 
result in sixth order accuracy for the phase error and eighth accuracy in 
the periodic energy. By generalizing the recently discovered
one-parameter family of fourth order symplectic algorithms\cite{chinchen02}, 
we can minimize the energy and phase error to even higher order.
The results of this study provides a direct verification of Chin's 
correctability criterion \cite{symcorr} for correcting a symplectic 
algorithm to higher order. In particular, we showed that the correctability
criterion is superior to the conventional wisdom of minimization 
of the sum of squares of error coefficients. The most important conclusion
of this work is that for periodic motion, the phase error is a more 
discriminating gauge of an algorithm's effectiveness than the energy error. 

   As a more important application of the phase error analysis, 
we track the orbital precession angle of the 2D Kepler problem by 
monitoring the rotation angle of the Laplace-Runge-Lenz vector\cite{hiord}. 
By comparing with various recent fourth order algorithms, we demonstrated
the uniqueness of forward symplectic algorithm in minimizing the 
phase error of this important class of celestial mechanics problems. 

\section*{Acknowledgements}
This work was supported in part, by a National Science Foundation
grant (to SAC) No. DMS-0310580.

\section*{Appendices}
\addcontentsline{toc}{part}{Appendices}

\appendix

\section{Fourth Order Error Coefficients}\la{4th1}

The error coefficients of the fourth order forward algorithm (\ref{algacop})
can be computed in terms of algorithm's factorization coefficients
via a Mathematica program\cite{lang4}. They are:
\bea
\t1 &=& 2\,(t_0+t_1)\,, \la{ct}\\
\v1 &=& (2\,v_1+v_2)\,, \la{cv}\\
\ct &=& -\fft16\left[\,t_1^2\,(-4\,v_1+v_2)+t_0^2\,(2\,v_1+v_2)
+2\,t_0\,t_1\,(2\,v_1+v_2)\,\right]\,, \la{t3}\\
\cv &=& \fft16\left[\,6\,u_0-t_0\,(2\,v_1+v_2)^2
+t_1\,(2\,v_1^2+2\,v_1\,v_2-v_2^2)\,\right]\,, \la{v3}\\
\tt &=& \fft1{360}\left[\,7\,t_0^3\,(t_0+4\,t_1)\,(2\,v_1+v_2)\right.\,\nn\\
&&\left.+\,t_1^3\,(4\,t_0+t_1)\,(7\,v_2-16\,v_1)
+6\,t_0^2\,t_1^2\,(4\,v_1+7\,v_2)\right]\,, \la{t50}\\
\tv &=& \fft1{90}\left[\,2\,t_0^2\,(t_0+3\,t_1)\,(2\,v_1+v_2)^2\,\right.\nn\\
&& \left.-6\,t_0\,t_1^2\,(6\,v_1^2+v_1\,v_2-v_2^2)
+t_1^3\,(8\,v_1^2-7\,v_1\,v_2+2\,v_2^2)\right]\,,\la{v50}\\
\vt &=& \fft1{60}\left[\,t_0^3\,(2\,v_1+v_2)^2\
+\,t_1^2\,\left(10\,(3\,\a-1)\,u_0
+t_1\,(-16\,v_1^2+4\,v_1\,v_2+v_2^2)\right)\right.\,\nn\\
&&+\,t_0^2\,\left(-10\,u_0+t_1\,(2\,v_1^2
+2\,v_1\,v_2+3\,v_2^2)\right)\,\la{t5}\\
&&\left.+\,t_0\,t_1\,\left(-20\,u_0
+t_1\,(12\,v_1^2+2\,v_1\,v_2+3\,v_2^2)\right)\right]\,,\nn\\
\vv &=& \fft1{60}\left[\,2\,t_0^2\,(2\,v_1+v_2)^3
-4\,t_0\,(2\,v_1+v_2)\,\left(5\,u_0
+t_1\,(v_1^2+v_1\,v_2-v_2^2)\right)\right.\,\la{v5}\\
&&\left.+\,t_1\,\left(10^{}\,u_0\,(2\,v_1+(3\,\a-2)\,v_2)
-\,t_1\,(4\,v_1^3+v_1^2\,v_2
+3\,v_1\,v_2^2-2\,v_2^3)\right)\right]\, .\nn
\eea
In order for the algorithm to be fourth order, we must have
$\t1=\v1=1$ and $\ct=\cv=0$. These four constraints can be satisfied
by
\be
t_1=t_2=\fft12-t_0\, ,\quad t_3=t_0\, ,\quad
v_1=v_3=\fft1{6\,(1-2\,t_0)^2}\, ,\la{cparam}
\ee
\be
v_2=1-(v_1+v_3)\, ,\quad
u_0 =\fft{1}{12}\left[1-\fft{1}{1-2\,t_0}+\fft{1}{6\,(1-2\,t_0)^3}\right]\, .
\\[.2cm]\la{cu0}
\ee
This is the family of fourth order algorithms (\ref{algacop})
with parameters $t_0$ and $\alpha$. For the harmonic oscillator,
$\tt$ and $\tv$ vanish identically. A fourth
order algorithm can be corrected to sixth order if one can set $\vt=\vv$.
Substituting (\ref{cparam}) and (\ref{cu0}) into (\ref{t5}) and (\ref{v5}),
gives $\vt$ and $\vv$ as functions of the parameters $t_0$ and $\a$,
\ie
\bea
\vt&=&\fft{1+5\,\a-12\,t_0\,(1+5\,\a+20\,\a\,t_0\,(-1+t_0))}
{2880\,(1-2\,t_0)}\, ,\la{ctcvalpha}\\[.3cm]
\vv&=&\fft{1+10\,\a-6\,t_0\,(3+30\,\a-t_0\,(9+210\,\a+8\,t_0\,
(1-85\,\a-3\,t_0\,(1-40\,\a+20\,\a\,t_0))))}
{4320\,(1-2\,t_0)^4}\, . \nn
\eea
Solving for $\vt=\vv$ determines $\alpha$ as a function of $t_0$:
\be
\a=\fft{1+6\,t_0\,(-3+4\,t_0\,(6+t_0\,(-23+24\,t_0)))}
{5\,(1-12\,t_0\,(1-2\,t_0)^2)\,(1-6\,t_0\,(1+2\,t_0-4\,t_0^2))}\, .
\la{alpha}
\ee
However, there exists no real solution of the parameters for which both,
$\vt$ and $\vv$ can be set to zero, \ie, we can have an
algorithm that is correctable to sixth order, but not a real sixth order
algorithm.



\newpage
\begin{thebibliography}{99}
\bibitem{wisdom} J. Wisdom and M. Holman, Astrophys. J., {\bf 102} (1991) 1528.
\bibitem{yoshi} H. Yoshida, Celest. Mech. {\bf 56} (1993) 27.
\bibitem{mcl95} R. I. McLachlan, SIAM J. Sci. Comput. {\bf 16}, 151 (1995).
\bibitem{cha96} P.J. Channell and F.R. Neri, F.R.,
        'An introduction to symplectic integrators', in
          {\it Integration algorithms and classical mechanics},
          (Toronto, ON, 1993), Fields Inst. Commun., 10,
          Amer. Math. Soc., Providence, RI, P.45-58.
\bibitem{mcl02} R. I. McLachlan and G. R. W. Quispel, Acta Numerica,
          {\bf 11}, 241 (2002).
\bibitem{bat99}R.H. Battin, {\it An Introduction to the
          Mathematics and Methods of Astrodynamics, Reviesed Edition}, AIAA, 1999.
\bibitem{shita} H. Kinoshita, H. Yoshida, and H. Nakai,
         Celest. Mech. {\bf 50} (1991) 59-71.
\bibitem{gladman} B. Gladman, M. Duncan and J. Candy,
         Celest. Mech. {\bf 52} (1991) 221.
\bibitem{cano} B. Cano and J.M. Sanz-Serna,
              SIAM J. Numer. Anal. {\bf 34} (1997) 1391.
\bibitem{ge}G. Zhong and J. E. Marsden,
            Phys. Lett. {\bf A133}, 134 (1988)
\bibitem{wis96}J. Wisdom, M. Holman, AND J. Touma, ``Symplectic correctors",
          in {\it Integration Algorithms and Classical Mechanics},
          J. E. Marsden, G. W. Patrick, and W. F. Shadwick, eds.,
          American Mathematical Society, Providence, RI, 1996.
\bibitem{mcl962}R. I. McLachan, ``More on symplectic correctors", in
        {\it Integration Algorithms and Classical Mechanics}, J. E. Marsden,
        G. W. Patrick, and W. F. Shadwick, eds.,
         American Mathematical Society, Providence, RI, 1996.
\bibitem{mar96}M. A. Lopez-Marcos, J. M. Sanz-Serna, and R. D. Skeel,
            in Numerical Analysis 1995, D. F. Griffiths and G. A. Watson,
            eds., Longman, Harlow, UK, 1996, pp. 107-122.
\bibitem{mar97}M. A. Lopez-Marcos, J. M. Sanz-Serna, and R. D. Skeel,
           SIAM J. Sci. Comput., {\bf 18} 223, (1997).
\bibitem{blan99}S. Blanes, F. Casas, and J. Ros,
        Siam J. Sci. Comput., {\bf 21}, 711 (1999).
\bibitem{symcorr}
           S. A. Chin, Phys. Rev. {\bf E 69}, (2004) 046118.
\bibitem{suss}G. J. Sussman and J. Wisdom with M. E. Mayer,
           {\it Structure and interpretation of classical mechanics }
                  MIT Press,Cambridge, Mass (2001).
\bibitem{suzfour}M. Suzuki, {\it Computer Simulation Studies in
            Condensed Matter Physics VIII},
           eds, D. Landau, K. Mon and H. Shuttler (Springler, Berlin, 1996).
\bibitem{chin} S.A. Chin, Physics Letters {\bf A226}, (1997) 344.
\bibitem{hiord} Siu A. Chin and Donald W. Kidwell, Phys. Rev. {\bf E 62}, (2000).
\bibitem{chinchen02}S. A. Chin and C. R. Chen,
                J. Chem. Phys. {\bf 117}, 1409 (2002).
\bibitem{chinchen03}S. A. Chin, and C. R. Chen,
         "Forward Symplectic Integrators for Solving Gravitational
          Few-Body Problems", arXiv, astro-ph/0304223, in press,
          Cele. Mech. Dyn. Astron.
\bibitem{forest} E. Forest and R. D. Ruth, Physica D {\bf 43} (1990) 105.
\bibitem{ome02}I. P. Omelyan, I. M. Mryglod and R. Folk,
               Phys. Rev. {\bf E66}, 026701 (2002).
\bibitem{ome03}I. P. Omelyan, I. M. Mryglod and R. Folk,
               Comput. Phys. Commun. {\bf 151} 272 (2003)
\bibitem{ti}M. Takahashi and M. Imada, J. Phys. Soc. Jpn {\bf 53}, 3765 (1984).
\bibitem{dragt} A. J. Dragt and J. M. Finn,
       J. Math. Phys. {\bf 17} 2215 (1976)
\bibitem{chinscuro}S. A. Chin and S. R. Scuro, ``Exact evolution of
       symplectic integrators and their phase error for the
       harmonic oscillator", arXiv math-phy/0408004.
\bibitem{sheng}Q. Sheng, IMA J. Num. Anaysis, {\bf 9}, 199 (1989).
\bibitem{suzukinogo}M. Suzuki, J. Math. Phys. {\bf 32}, 400 (1991).
\bibitem{goldman}D. Goldman and T. J. Kaper, SIAM J. Numer. Anal.,{ \bf 33},
                   349 (1996).
\bibitem{lang4} Harald A Forbert and Siu A Chin Phys. Rev.
                 {\bf E 63}, 016703 (2001).
\bibitem{dmc4} Harald A Forbert and Siu A Chin, Phys. Rev. {\bf B 63},
                 144518 (2001).
\bibitem{serna} J. M. Sanz-Serna and A. Portillo,
      J. Chem. Phys. {\bf 104}, 2349 (1996).


\end{thebibliography}
\end{document}